\documentclass[twocolumn,preprintnumbers,amsmath,amssymb,floatfix]{revtex4-1}

\usepackage{graphicx}
\usepackage{color}

\usepackage[linktocpage,bookmarksopen,bookmarksnumbered]{hyperref}
\usepackage{dcolumn}

\usepackage{amsmath,graphics,epsfig,color,verbatim,ulem}

\newcommand{\be}{\begin{equation}}
\newcommand{\ben}{\begin{equation*}}
\newcommand{\ee}{\end{equation}}
\newcommand{\een}{\end{equation*}}
\newcommand{\bs}{\begin{split}}
\newcommand{\es}{\end{split}}
\newcommand{\bmx}{\begin{array}}
\newcommand{\emx}{\end{array}}
\newcommand{\bea}{\begin{eqnarray}}
\newcommand{\bean}{\begin{eqnarray*}}
\newcommand{\eea}{\end{eqnarray}}
\newcommand{\eean}{\end{eqnarray*}}
\newcommand{\dg}{^{\dagger}}
\newcommand{\dn}{^{\vphantom{\dagger}}}

\newcommand{\eps}{\epsilon}

\newcommand{\pref}[1]{(\ref{#1})}

\newcommand{\intob}[1]{\int_{0}^{\beta}{#1}}

\newcommand{\abs}[1]{\left\vert #1 \right\vert}

\newcommand{\ket}[1]{\left\vert #1\right\rangle}

\newcommand{\braket}[1]{\left\langle #1\right\rangle}

\newcommand{\mat}[1]{\left(\bmx{cc}#1\emx\right)}

\begin{document}

\setlength{\pdfpageheight}{\paperheight}
\setlength{\pdfpagewidth}{\paperwidth}


\title{Even-Odd Effect of a Spin-S impurity Coupled to a Quantum Critical System}


\author{Kun Chen}
\author{Yashar Komijani}

\affiliation{Department of Physics and Astronomy, Rutgers University, Piscataway, New Jersey 08854, USA}
\date{\today}


\begin{abstract}
{ We discuss an even-odd effect for an impurity with an $N$-fold degenerate internal states immersed in a two-dimensional superfluid--Mott-insulator quantum critical bath which is described by a spin-$S$ XY Bose-Kondo impurity model with $N=2S+1$. Using a dimensional- and momentum-cut off regularized renormalization group and unbiased large-scale Monte Carlo numerical simulations, we establish the phase diagram for the $S=1$ impurity with all relevant terms included. We show that a 3-fold degenerate $S=1$ impurity coupled to a critical bath is fully screened, in qualitative contrast to the spin-1/2 case where the impurity is only partially screened. We then argue that all impurities with odd 2$S$ degeneracy share the same universal physics as the spin-1/2 case, and all impurities with even 2$S$ degeneracy are as the spin-1 case. We validate our conjecture with unbiased Monte Carlo simulations up to $S=2$. A physical consequence of this even-odd effect is that two spin-$1/2$ XY Bose-Kondo impurities---or two halons in the bosonic language---in the critical bath form a long-range entangled state at a sufficiently low temperature, which can be realized in ultracold atoms in an optical lattice.}
\end{abstract}

\maketitle

{\it Introduction --} The recent years have witnessed a widespread increase of interest in impurities coupled to a bosonic bath. The impurity immersed in a weakly and strongly interacting bosonic bath are shown to exhibit many exotic emergent phenomena (e.g., the Bose polaron~\cite{compagno2017tunable,grusdt2017strong}, the halon effect~\cite{Punk13, huang2016trapping,Whitsitt17,chen2018halon} and the trapping collapse effect~\cite{chen2018trapping}). Many of those impurity physics are directly accessible with modern techniques in ultracold atoms experiments \cite{Greiner02,jorgensen2016observation, chen2018halon}. Furthermore, the bosonic impurity also finds applications in a numerical method: a framework of dynamical mean field theory is recently developed on top of a bosonic impurity model, providing a new powerful tool to attack strongly interacting bosonic systems~\cite{hu2009dynamical,byczuk2008correlated,anders2011dynamical}.


It has been recently demonstrated \cite{huang2016trapping,chen2018halon} that a sufficiently strong trapping potential in a Bose-Hubbard system, tuned to its bulk { superfluid--Mott-insulator transition leads to a boundary critical phenomena with XY universality class}. 

The bulk is described by the 2+1 dimensional O(2) $\phi^4$ theory
\be
S_{bath}=\int{d^{d+1}}x\Big[(\partial^\mu\vec \phi)(\partial^\mu\vec \phi)+\frac{r}{2}\vec\phi^2+\frac{g_0}{4!}(\vec\phi^2)^2\Big]
\ee
where $\vec \phi=(\phi_x,\phi_y)$ and the parameter $r$ is tuned to the critical point which is described by the Wilson-Fisher (WF) finite-coupling fixed point \cite{Fisher89,Sachdev}. The boundary critical point induced by the trapping potential is then described by the effective model
\be
{H}_{BK}={H}_{bath}[{\phi}^-, {\phi}^+]+\gamma[{\phi}^-(0){S}^++h.c],\label{eqBK}
\ee
where $\phi^\pm=\phi_x\pm i\phi_y$ and in the simplest case $S^+$ transforms under the fundamental spin-1/2 representation of SU(2) group. A (pseudo) spin-S magnetic impurity coupled to a bosonic bath is generally called the Bose-Kondo model \cite{Sengupta00,Smith99}. The SU(2) version of { this} problem arises by doping an antiferromagnet at its quantum critical point \cite{Sachdev99,Vojta00,Sachdev01,Punk13}. Here, we are intersted in the XY Bose-Kondo model \cite{Whitsitt17,chen2018halon,Zarand02}.

The special case $S=1/2$ is argued to feature a boundary quantum critical point (BQCP) with $S=1/2$ XY Bose-Kondo universality class \cite{huang2016trapping,Whitsitt17,chen2018halon}. The hallmark of this BQCP is the partial screening of the impurity which leads to the so-called {\it halon effect}: When a small polarization field $h_z{S}_z$ is used to lift the two-fold degeneracy of the ground state, the half-integer projection of the pseudo-spin on its z-axis gets delocalized into a halo of critically divergent radius $\sim 1/|h_z|^{\nu_z}$ where $\nu_z=2.33(5)$~\cite{huang2016trapping}. In other words, the integer charge carried by  { a trapping potential impurity}, gets fractionalized into two parts: a microscopic core with half-integer charge and a critically large halo carrying a complementary charge of $\pm 1/2$. {This critical impurity state---refered to as halon---describes the emergent physics of { one potential scattering impurity} in ultracold atoms in optical lattice near the superfluid--Mott-insulator quantum critical point \cite{chen2018halon}}.

{ In experiments, one can apply multiple potential scattering impurities in the system. Tuning the potential strength then leads to multi-halon---or equivalently, multiple spin-$1/2$ XY Bose-Kondo impurities---coupled to the same critical bath. The ground state of such system is still an open question. We find that this problem is closely related to the problem of the XY Bose-Kondo impurity problem with a generic $S>1/2$. Indeed, in the long wavelength limit, the problem of two spin-$1/2$ impurities is effectively described by a spin-$1$ Bose-Kondo model \cite{SM} with 3-fold degeneracy (up to a small perturbation). This motivates us to ask the question: what is the ground state by coupling a $(2S+1)$-degenerate spin impurity to a quantum critical bath as in Eq. (\ref{eqBK})?  


Whitsitt and Sachdev have recently studied the boundary critical point for the spin-$S$ XY Bose Kondo model in an $O(N)$ critical bosonic bath \cite{Whitsitt17}. For $S >1/2$ and $N=2$ symmetry, the impurity models admit (more and more) relevant operators that need to be fine-tuned to maintain the $(2S+1)$-degeneracy of the ground state. However, there exist also stable quantum phases with partial lifting of $(2S+1)$-degeneracy, which is perhaps more relevant to the experiments.
}

{ Here, we argue that the ground state of a spin-$S$ impurity coupled to a quantum critical bath exhibits an even-odd effect: for all even $2S$ impurity, the degeneracy is completely lifted and the impurity ground state has a sharply defined charge quantum number; while for all odd $2S$, the impurity is partially screened and share the same universal physics as the $S=1/2$ case. }

\begin{figure}[tp!]
\includegraphics[width=1\linewidth]{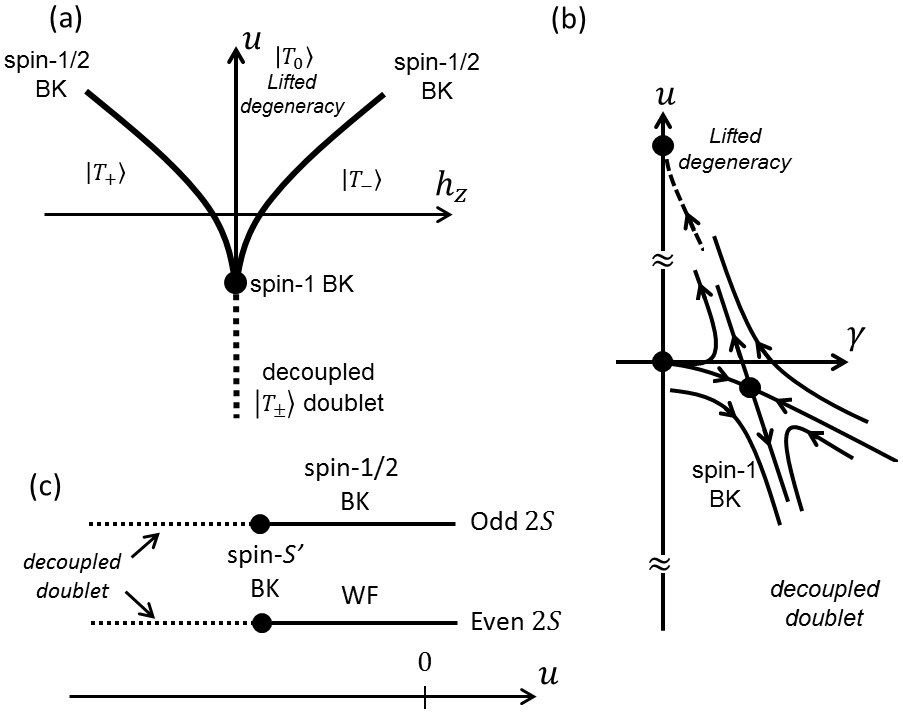}
\caption{\small (a) The phase diagram of the spin-1 with a XY coupling to a critical bosonic bath is very similar to the decoupled impurity. $\ket{T_0}$ corresponds to full screening, whereas spin-1 BK represents partially screened case. The dashed line represents a level-crossing of the decoupled doublet. (b) The perturbative RG flow for $h_z=0$ (solid line) and the speculative extrapolation (dashed line) show that fixed point requires fine tuning $u$. (c) The phase diagram at $h_z=0$ for a general spin-$S$ impurity.
}\label{fig0}
\end{figure}

To obtain this result, we first establish the complete phase diagram, as shown in Fig.\,\pref{fig0}, of the Spin-$1$ Bose-Kondo model with all relevant terms included. The transition lines and the nontrivial critical exponents are calculated and cross-checked with an $\epsilon$-expansion renormalization group (RG) approach and unbiased large-scale Quantum Monte Carlo (QMC) numerical simulations. Our calculations show that the $N=3$ degeneracy of the impurity as shown in Eq. (\ref{eqBK}) is effectively lifted by the impurity-bath interaction, which causes the pseudo-spin to be fully screened in the long-wave-length limit, in qualitative contrast to the previously established spin-1/2 impurity. 

We then use symmetry arguments to show that all $2S$ odd impurities share the same physics as the spin-1/2 Bose-Kondo impurity, whereas all $2S$ even impurities are different and behave as the spin-1 impurity. This conjecture is validated with the unbiased Monte Carlo simulations up to $S=2$.

{ The even-odd effect implies that two spin-1/2 impurities form a long-range entangled state regardless of their distance $d$ (note that the bulk system is at a quantum critical point with infinitely large correlation length). In the bosonic language, two halons are paired by sharing one boson. This prediction can be tested in the future experiment with ultracold atoms in an optical lattice.
}

This scenario can be compared to the two-impurity spin-1/2 Fermi-Kondo problem \cite{Jayaprakash81,Jones88,Affleck92,Affleck95}.
The RKKY interaction $J_H\vec S_1\cdot\vec S_2$ between the impurities competes with the local Kondo screening $J_K\vec S_i\cdot\vec s_i$ by the conduction band spin density $\vec s_i$. When $J_H$ is large and positive, the two spins form a singlet and the conduction electrons are reflected with the phase shift $\delta_c=0$. In the opposite regime of large negative $J_H$, the two impurities form a spin-1 impurity which is screened by two conduction electron bands with the phase shift $\delta_c=\pi$. Since the phase shift has to be either $0$ or $\pi$ in presence of particle-hole symmetry, there has to be a quantum phase transition at which the phase shift jumps. The (single-) Bose-type of coupling in Eq.\,\pref{eqBK} and the XY symmetry change this picture qualitatively.


{\it Spin-$1$ Bose-Kondo model -- } We consider the spin-$1$ Bose Kondo model
\be
{H}_{BK}={H}_{bath}[{\phi}, {\phi}^\dag]+\gamma[{\phi}(0){S}^\dag+h.c]+u({S}_z)^2.\label{eqBK1}
\ee

This is the most general form of the interaction with $Z_2$ and $U(1)$ invariance,
\bea
\hspace{-.5cm}U(1)&:&\quad \phi^\pm\to e^{\pm i\alpha}\phi^\pm, \quad S_\pm\to e^{\pm i\alpha}S^\pm, \quad S_z\to S_z\\
Z_2&:&\quad \phi^\pm\to \pm i\phi_\mp, \quad S_\pm\to \pm i S_\mp, \quad S_z\to -S_z.
\eea
which are subgroups of the SU(2) group of the spin. The problem without the $u$-term has been studied in \cite{Whitsitt17} using RG and $\eps$-expansion. Indeed in presence of SU(2) symmetry, only (functions of) $\vec S^2$ can appear which are trivial as $\vec S^2$ is a conserved quantity. However, in a spin-1 system, the $u$-term is allowed by both $U(1)$ and $Z_2$ symmetries and is dynamically generated at low energies.

{ For the spin-impurities coupled to interacting bath, we characterize the screening based on the fluctuations of the total charge in the system $Q=\sum_i \phi\dg_i \phi\dn_i+S_z$. At low-energies, the residual charge fluctuations $\braket{\delta Q^2}-\braket{\delta Q^2}_{\rm bulk}$ determine if the impurity is fully or partially screened or it has decoupled from the bath.}

The quadratic part of the bath action has the zero temperature equal-position two-point function
\be
\hspace{-.25cm}D_d(\tau)\equiv\braket{T\phi_\alpha(\tau)\phi_\alpha}\sim\frac{1}{\abs{\tau}^{d-1}}\label{eqD}
\ee
for $\alpha=x,y$. At $d=3-\eps$, the bulk is weakly interacting. Additionally, the impurity-bath coupling in Eq.\,\pref{eqBK1} is barely relevant. 
This allows us to treat the impurity problem in Eq.\,\pref{eqBK1} perturbatively and then extrapolate the result to $\eps\to 1$. 

First we discuss the effect of the $u$-term in lifting the internal degeneracy of the triplets in the spin-1 impurity. At large positive values $u\to\infty$, the impurity has a non-degenerate ground state $\ket{T_0}$, with $\braket{S_z^2}_{T\to 0}=0$. The single-boson coupling is thus projected to zero and the second order perturbation theory leads to 
\be
H_{imp}\to \frac{\gamma^2}{u}[\phi^+(0)\phi^-(0)+h.c.]
\ee
which is a potential scattering with the strength $\gamma^2/u$. For $\gamma^2/u\ll 1$ this is irrelevant at $d=3-\eps$, including the marginal case of $d=2$. \cite{chen2018halon}.

In the opposite regime of $u\to-\infty$, the $\ket{T_+}$ and $\ket{T_-}$ form a degenerate ground state with $\braket{S_z^2}_{T\to 0}=1$, which act as an effective spin-1/2 but their mixing with the bosons require two-boson exchange:
\be
H\to \frac{\gamma^2}{u}[(\phi^-)^2(S^+)^2+h.c.].
\ee
This is again irrelevant for $d=3-\eps$ and $\eps\ll 1$, but becomes marginal at $d=2$. There is no reliable way to study this term in $d=2$, since the bulk is strongly interacting. However, by a crude replacement of the bulk with a non-interacting bath of bosons with the propagator $D_d(\tau)\sim 1/\tau^{d-1}$, we can study this term in $d=2-\eps$. Second order perturbation theory contains $D_{2-\eps}^2(\tau)\sim 1/\tau^{2-2\eps}$ instead of $D_{3-\eps}(\tau)\sim 1/\tau^{2-\eps}$. The only modification to the beta function of spin-1/2 BK \cite{Whitsitt17} is to replace $\eps\to 2\eps$. This gives a non-trivial fixed point that goes to zero as $\eps\to 0$. We conclude that for non-interacting bosonic bath this term is irrelevant. In presence of an interacting bath, we have to rely on our QMC simulations. Indeed, we will see that for $u<u_*$, the system flows to a WF fixed point and a decoupled doublet. We have also confirmed that this interaction is irrelevant in an effective spin-1/2 model. 

Since the $u$-term is relevant, the boundary critical point requires fine tuning of at least one parameter and this is expected to affect the exponents
\be
\hspace{-.4cm}\braket{T_\tau S_z(\tau)S_z}\sim\frac{1}{\abs{\tau}^{\eta_z}}, \qquad {\small{\sum_\alpha}}\braket{T_\tau S_\alpha(\tau)S_\alpha}\sim\frac{1}{\abs{\tau}^{\eta_\perp}},
\ee
at the critical point. This is plausible considering that the ${\cal O}(\eps^2)$ exponents computed in \cite{Whitsitt17} for $S=1$ and the $O(2)$ model, which are $\eta_\perp:\eps-5.2838\eps^2$ and $\eta_z:2\eps-8.5676\eps^2$ become negative for $\eps>0.23$. These exponents remain unchanged to order ${\cal O}(\eps)$ in presence of the $u$; $\eta_\perp=\eps$ and $\eta_z=2\eps$; Moreover, we can compute the exponent corresponding to the relevant operator
\be
\braket{T_\tau S_z^2(\tau)S_z^2}\sim\frac{1}{\abs{\tau}^{\eta_u}},
\ee
which we compare with the Monte Carlo. \\

{\it RG analysis --} The $uS_z^2$ term leads to IR divergences in perturbation theory which are cut by the temperature, i.e. expanding in terms of $u\beta$ instead of $u$. For sufficiently small $u$, $\beta$ can be very large and we use the zero-temperature form\,\pref{eqD} of the boson propagator.

Additionally, the dynamic generation of mass means that perturbation theory is plagued with UV divergences that are not cured by the dimensional regularization. Hence, in addition to dimensional regularization, we introduce a momentum cut-off $\mu_0$. Eventually, $\beta$ is replaced with the energy $\mu$ of interest and $\mu_0$ is sent to infinity after renormalization. 

Renormalization is achieved by introducing scale-dependent operators and coupling-constants \cite{ZinnJustin}. In addition to $\phi\to\sqrt Z\tilde\phi$ in the bulk, we have \cite{Whitsitt17}
\be
S_z\to \sqrt{Z_z}\tilde S_z, \quad S_{x,y}\to \sqrt{Z_\perp}\tilde S_{x,y}, \quad S_z^2\to Z_u\tilde{S_z^2}. 
\ee
Note that $S_z^2$ is a distinct operator and gets its own renormalization factor. We find that the correlation functions of the renormalized operators are independent of the UV scale $\mu_0$ and analytical in $\eps$ \cite{SM}, provided that they are expressed in terms of renormalized couplings constants and we choose $Z$ parameters to absorb the poles in $\eps$. 
The details are discussed in supplementary material \cite{SM}. We find
\bea
\frac{d\gamma}{d\ell}&=&\frac{\gamma}{2}[\eps-\gamma^2(1+u)]\\
\frac{du}{d\ell}&=&u+(1+3u)\gamma^2\label{eqbeta}
\eea
where $d\ell\equiv-d\log\mu$. Fig.\,\ref{fig0}(b) shows the RG flow in the vicinity of the fixed point. To ${\cal O}(\eps)$ the factors $1+u$ in the first line and $1+3u$ in the second line, can be neglected. These equations have a non-trivial fixed point at $(\gamma_*^2,u_*)=({\eps},-\eps)$. Since $\eps\to1$, $\gamma_*^2\to 1$ and $u_*\to -1$. At the QCP the critical exponents of spin susceptibility $\chi_\alpha=\int{d\tau}\braket{TS_\alpha(\tau)S_\alpha}\propto T^{3-2/\nu_\alpha}$ remain unchanged. But we can compute one more exponent corresponding to $\chi_u=\int{d\tau}\braket{TS_z^2(\tau)S_z^2}$. To ${\cal O}(\eps)$, this is given by
\be
\eta_u=-\frac{d\log Z_u^2}{d\log\mu}=-2\frac{d\log Z_u}{d\gamma}\beta_\gamma=6\gamma^2=6\eps
\ee
from which we obtain $\nu_u\equiv(1-\eta_u/2)^{-1}=1+3\eps\to 4$.\\

\begin{figure}[tp!]
\includegraphics[width=1.0\linewidth]{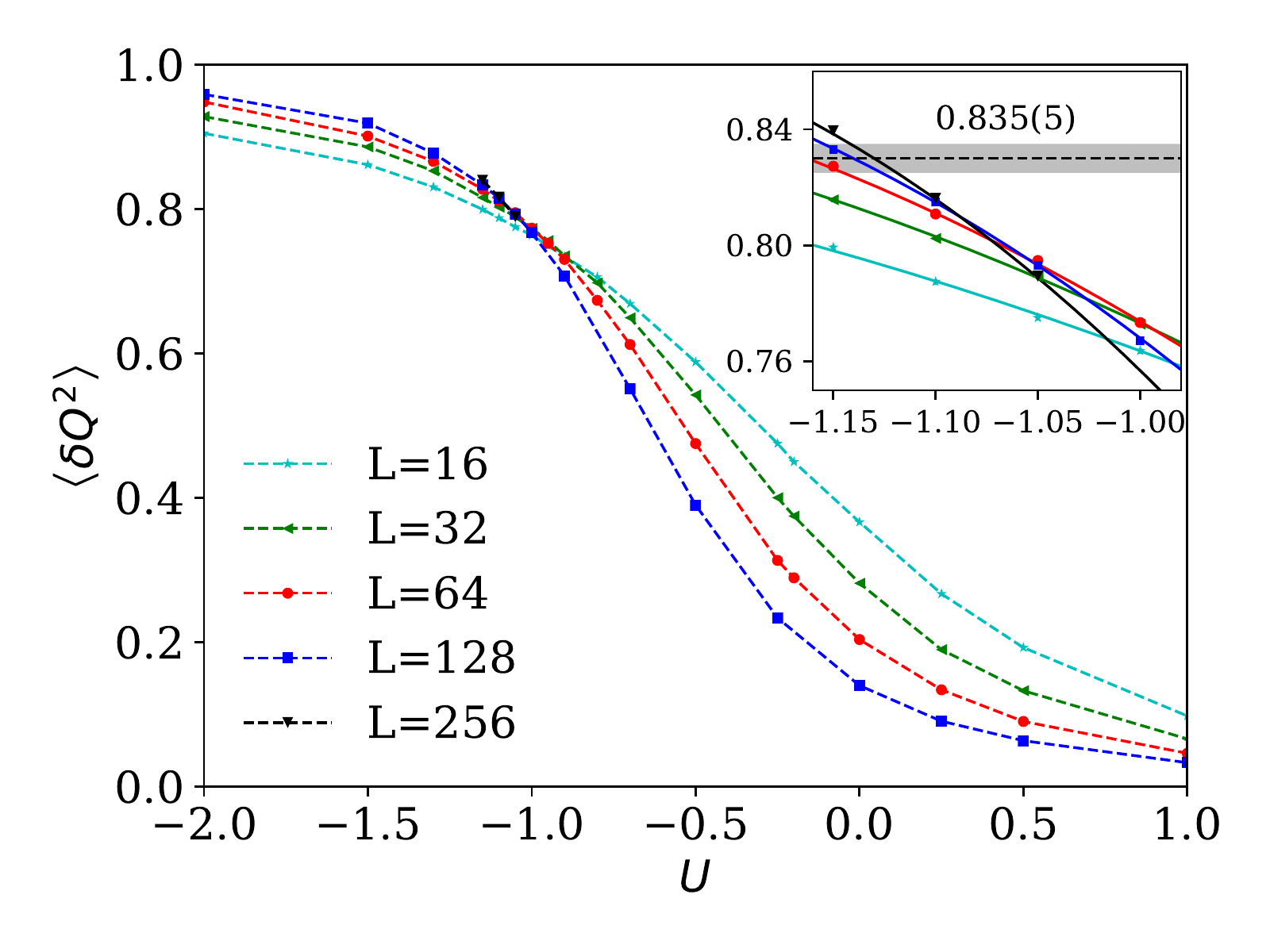}
\caption{The change of the total charge number fluctuation $\left< \delta Q^2 \right>=\left< Q^2 \right>-\left<Q \right>^2$ (subtracting the universal bulk contribution $\left< \delta Q^2 \right>_0=0.5160(6)$~\cite{chen2014universal}) as a function of the impurity interaction strength $U$. The inset shows that the curves near the critical strength $U_c=1.12(2)$ can be fitted with the same scaling ansatz $q_0+q_1(U-U_c)L^{1/\nu_u}+q_2(U-U_c)^2L^{2/\nu_u}+c/L^\omega$. We extract the universal constant $q_0=0.835(5)$ and the boundary critical exponent $\nu_u=3.6(3)$.
}
\label{fig1}
\end{figure} 

{\it Numerical Analysis -- } {We now study this impurity problem with an effective lattice model, which allows an efficiently unbiased Monte Carlo simulation using the worm algorithm. The details and the definition of this lattice model can be found in the supplementary material \cite{SM}. Here we show the main results obtained by large scale simulations and finite size scaling analysis. 

Fig.\,\pref{fig1} shows the charge fluctuations $\braket{\delta Q^2}$ as a function of the relevant term $u$ for various system sizes.  For each curve with a given system size, a bulk WF value of $\braket{\delta Q^2}_0\sim 0.516$ is subtracted from the data. A boundary quantum critical point which features the partially screened fixed point is found at $u_*=-1.12(2)$. For $u>u_*$ (including $u=0$) the system flows to WF fixed point, as if the system has fully screened the impurity.  Whereas for $u<u_*$ it flows to a WF and a decouple doublet with $\braket{\delta Q^2}-\braket{\delta Q^2}_0\rightarrow 1$. At $u=u_*$, a universal charge fluctuation of $\braket{\delta Q^2}-\braket{\delta Q^2}_0=0.835(2)$ and a critical exponent $\nu_u=3.6(3)$ is extracted, in good agreement with $\nu_u\sim 4$ from RG.
}

\begin{figure}[tp!]
\includegraphics[width=1.0\linewidth]{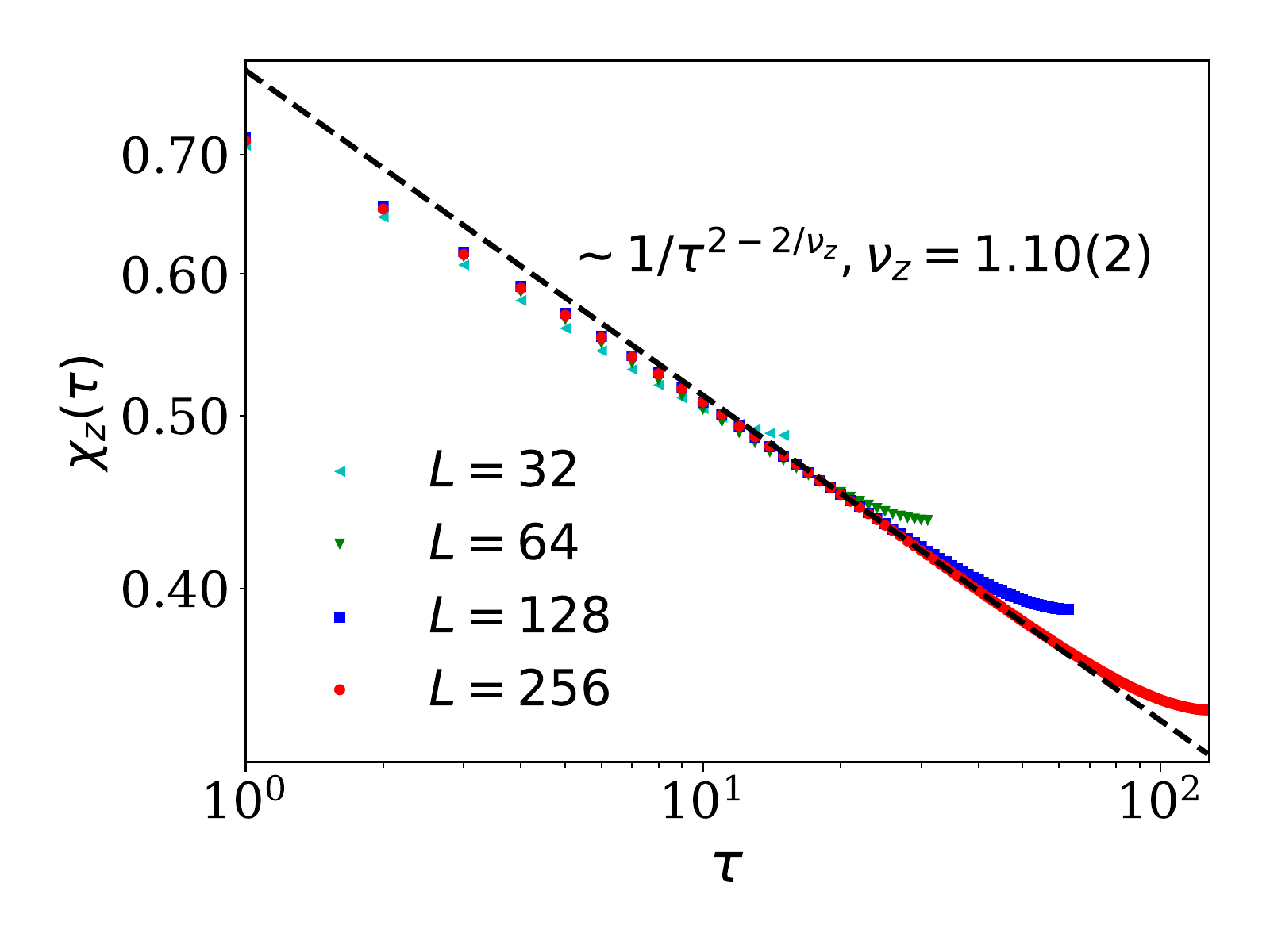}
\caption{\small The longitudinal spin-spin correlation function in the imaginary time $\chi_z(\tau)=\left< {S}^z(\tau){S^z}(0)\right>$ at the BQCP for different system sizes $L$. In the thermodynamic limit $L\rightarrow \infty$, there develops a power-law tail (the dashed line) expected from the general analysis (see the text), and yielding $\nu_z=1.10(2)$.
}
\label{fig2}
\end{figure}

Fig.\,\pref{fig2} shows the $\chi_z(\tau)=\braket{S_z(\tau)S_z}$ correlation function as a function of $\tau$ for various system sizes. $\chi_z(\tau)$ becomes a power-law with the exponent $\nu_z=1.10(2)$, corresponding to $\eta_z=0.18$ in marked contrast to the leading value of $\eta_z\sim 2\eps$ from RG analysis.

{\it Even-odd Effect -- } We showed that in the spin-1 case, one needs to fine tune the coefficient of $S_z^2$ in order to flow to the critical point. In impurities with higher spin-representations,  more independent operators appear that lift various degeneracy. However, we can argue that even $2S$ and odd $2S$ behave qualitatively different. In the former case, the ground state is generically non-degenerate or degenerate but with large spin-differences and therefore, the impurity decouples and bulk remains at the WF fixed point. Whereas, odd $2S$ in presence of $Z_2$ symmetry, is guaranteed to have an (at least doubly) degenerate ground state. Moreover, the sign of the dynamically generated $u$-term can be shown to be positive, and therefore, the doublet are $\Delta m_z=1$ spin apart and are capable of coupling to the bath. 


\begin{figure}[tp!]
\includegraphics[width=1.0\linewidth]{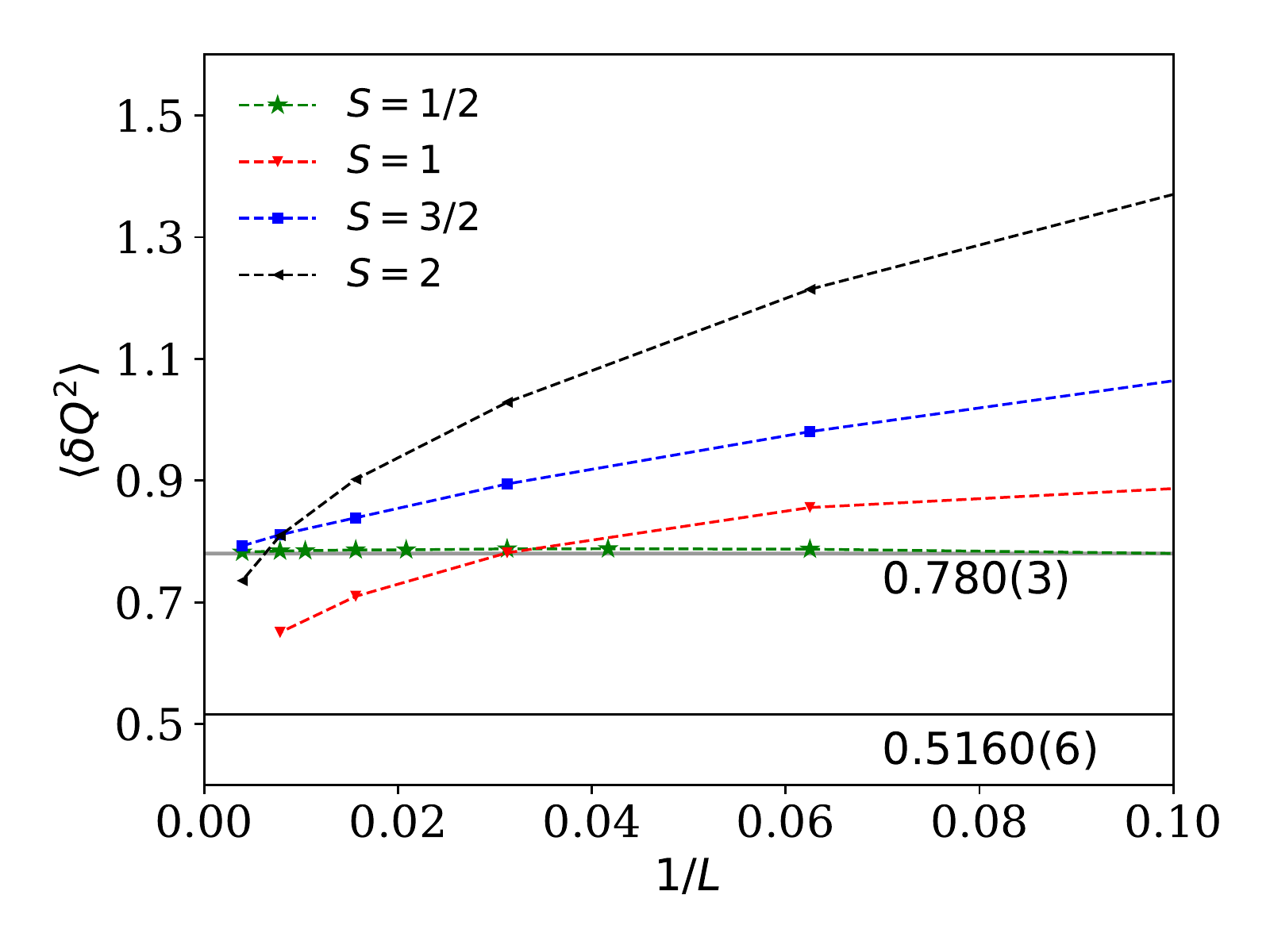}
\caption{\small The finite-size flows of the total charge number fluctuation $\left< \delta Q^2 \right>=\left< Q^2 \right>-\left<Q \right>^2$ for an impurity with spin $S=1/2,1,3/2$ and $2$. In the thermodynamic limit $1/L \rightarrow 0$, an impurity with even spin and odd spin demonstrates different trends. An even-spin impurity is fully screened, because it flows to the bulk universal constant $0.5160(6)$~\cite{chen2014universal}, which characterizes the Wilson-Fisher universality class. On the other hand, An odd-spin impurity is only partially screened, and it flows to a different universal constant $0.780(3)$, which characterizes the spin-$1/2$ XY Bose Kondo boundary universality class.
}
\label{fig3}
\end{figure} 

Fig.\,\pref{fig3} shows QMC results on charge fluctuations for different spin-$S$ impurities as a function of system size. Only XY coupling is included in this calculations and no fine-tuning. For $S=1$ and $S=2$, the charge fluctuations approach the WF fixed point at low-energies, indicating a full screening. However, $S=1/2$ and $S=3/2$ have the same IR fixed point, higher than the WF value, indicating a partially screened impurity. This is in agreement with the above argument and the phase diagram suggested before.

{\it Discussion --}{ { In the multi-impurity problem, the even-odd effect indicates that two (or any even number of) spin-$1/2$ impurities form a entangled state with $S_z=0$ projection at a sufficiently low temperature. In the language of two potentials forming two halons, each halon provides $1/2$ boson which is delocalized in the environment. Naturally, they merge into one shared boson by the two potentials. The shared boson induces an attractive interaction that decays as $\sim 1/d$ where $d$ is the distance between two potentials. If the halons are mobile and have sufficiently large mass, this attractive interaction can potentially give rise to a bound state. Such physics can be realized in ultracold atoms in optical lattice; one may create two halons with two tightly focused laser beams which are perpendicular to the 2D lattice and with critical strength. At a sufficiently low temperature, the particle number at two impurity sites are anti-correlated against any low-energy and long-wave-length probe. For example, if one impurity traps more charge by adibatically increasing the laser strength, the charge around the other impurity decreases to keep the total charge a good quantum number. }
}

In conclusion, we have studied spin-$S$ impurities coupled via an XY coupling to a critical bosonic bath. The simplest case $S=1$ admits relevant symmetry-preserving terms that need to be fine-tuned at the criticality. We have computed the corresponding critical exponents which have good agreement with QMC. We have also argued that impurities with $2S$ even and $2S$ odd behave qualitatively different and demonstrated this using QMC. { We also propose the experimental protocol to observe this effect in ultracold atoms in an optical lattice.}

\textbf{Acknowledgments:} 
K.C. and Y.K. contributed equally to this work. {It is a pleasure to thank Boris Svistunov for multiple discussions, his critical reading and detailed comments that helped to improve the manuscript.} The authors also acknowledge valuable discussions with Nikolay Prokof'ev and Youjin Deng. 

\bibliography{even-odd}{}

\newpage

\section*{Supplementary Material}
\subsection{Relation with two-impurity problem}
The problem of two spin-1/2 impurities with $U(1)$ symmetry is described by the Hamiltonian
\bea
H&=&H_{bath}+\sum_i\gamma[\phi^+(\vec r_i)S_i^-+h.c.]\\
&&+J_H(S_1^+S_2^-+h.c.)
\eea
where $\abs{\vec r_2-\vec r_1}=d$. At the $d=3-\eps$, the $\gamma$-couplings are barely relevant. At long wavelength (energies small compared to $1/d$) we have
\be
\phi^\pm(\vec r_1)\approx \phi^\pm(\vec r_2) \to \phi^\pm (\bar{\vec r})
\ee
as the corrections are highly irrelevant. Here, $\bar{\vec r}=(\vec r_1+\vec r_2)/2$. Therefore, the effective action becomes
\bea
H&=&H_{bath}+\gamma[\phi^+(\bar{\vec r})(S_1^-+S_2^-)+h.c.]\\
&&+J_H(S_1^+S_2^-+h.c.).
\eea
The two spins can be written in terms of a singlet and a triplet and the former has decoupled from the bosons. Therefore, up the a constant the Hamiltonian is equal to the spin-1 Hamiltonian of Eq.\,(6) where $S_1^\pm+S_2^\pm\to S^\pm$.

\subsection{Perturbation theory}
In this section we calculate the correlation functions 
\bea
{\cal M}_\perp(\tau)&\equiv&\Sigma_\alpha\braket{T_\tau S_\alpha(\tau)S_\alpha(0)}\\
{\cal M}_z(\tau)&\equiv&\braket{T_\tau S_z(\tau)S_z(0)}\\
{\cal M}_{u}(\tau)&\equiv&\braket{T_\tau S_z^2(\tau)S_z^2(0)}-\braket{S_z^2}^2
\eea
using perturbation theory in $\gamma_0$ and $u_0$. Since $u={\cal O}(\eps)$ and $\gamma={\cal O}(\sqrt\eps)$ we stop at order ${\cal O}(\eps^2)$. We discuss the derivation for the first correlation functions and for the other two, only provide the final result.

By expanding to second order in $\gamma$ (but exact in $u$) we have
\bea
{\cal M}_\perp&=&{\cal M}_\perp^0+\frac{\gamma^2}{2}\int{d\tau_1d\tau_2}D_d(\tau_1-\tau_2)\Lambda_\perp(\tau,0;\tau_1,\tau_2)\nonumber
\eea
Here, $D_d(\tau)$ is the two-point function of the bosonic bath and is given by
\be
\hspace{-.25cm}D_d(\tau)\equiv\braket{T\phi_\alpha(\tau)\phi_\alpha}=\int{\frac{d^dkd\omega}{(2\pi)^{d+1}}}\frac{e^{-i\omega\tau}}{k^2+\omega^2}=\frac{\tilde S_{d+1}}{\abs{\tau}^{d-1}}\ee
where
\be
\tilde S_{d}=\frac{\Gamma(d/2-1)}{4\pi^{d/2}}.
\ee
Using $S_{\pm}(\tau)=e^{\tau H}S_{\pm}e^{-\tau H}$, the effect of $u$ can be calculated non-perturbatively.  The zeroth order in $\gamma$ is
\be
{\cal M}_\perp^0(\tau)=\frac{e^{-u\tau}+e^{-u(\beta-\tau)}}{1+2e^{-\beta u}}
\ee
For the order $\gamma^2$ we have
\bea
\Lambda_\perp&=&\sum_{\alpha\mu=x,y}
\braket{T_\tau S_\alpha(\tau)S_\alpha(0)\Big[S_\mu(\tau_1)S_\mu(\tau_2)-\braket{\dots}_u\Big]}_u\nonumber
\eea
The spins do not obey Wick's theorem and one has to calculate the trace over spins for each time-ordering separately \cite{Barzykin96,Whitsitt17}. Also, note that $S_\alpha$ is the generator of a spin-1 representation of SU(2). Therefore, $S_\alpha^2\neq 1$. For the case of ${\cal M}_\perp$ we have
\bean
\Lambda_\perp&=&\frac{2}{1+2e^{-\beta u}}\Big[2e^{-u(\tau+\Delta\tau)}+e^{-\beta u}e^{u(\tau+\Delta\tau)}\Big]\Big(\theta_1+\theta_2\Big)\\
&+&\frac{2}{1+2e^{-\beta u}}\Big[e^{-(\tau-\Delta\tau)u}+2e^{-\beta u}e^{u(\tau-\Delta\tau)}\Big]\theta_3\\
&+&\frac{2}{1+2e^{-\beta u}}\Big[e^{(\tau-\Delta\tau)u}+2e^{-\beta u}e^{-u(\tau-\Delta\tau)}\Big]\theta_4\\
&+&\frac{2}{1+2e^{-\beta u}}\Big[e^{-u\tau}e^{u(\tau_1+\tau_2)}+e^{u(\tau-\beta)}e^{-u(\tau_1+\tau_2)}\Big]\theta_5\\
&+&\frac{2}{1+2e^{-\beta u}}\Big[e^{u\tau}e^{-u(\tau_1+\tau_2)}+e^{-\beta u}e^{-u\tau}e^{u(\tau_1+\tau_2)}\Big]\theta_6\\
&-&4\frac{e^{-u\tau}+e^{-u(\beta-\tau)}}{(1+2e^{-\beta u})^2}\Big[e^{-u\Delta\tau}+e^{-u(\beta-\Delta\tau)}\Big]\theta_0,
\eean
where we defined $\Delta\tau\equiv \tau_1-\tau_2$ and $2\bar\tau=\tau_1+\tau_2$. Expansion of these expressions to first order in $u$ gives the terms ${\cal O}(\gamma^2)$ and ${\cal O}(u\gamma^2)$ that we are interested in here. The $\theta_i$ factors are short-hand notations for the following Heaviside functions:
\bea
&&\theta_1\equiv \theta(\tau_1>\tau_2>
\tau>0)\\
&&\theta_2\equiv \theta(\tau>0>
\tau_1>\tau_2)\\
&&\theta_3\equiv \theta(\tau>0>
\tau_1>\tau_2)\\
&&\theta_4\equiv \theta(\tau_1>\tau>
0>\tau_2)\\
&&\theta_5\equiv \theta(\tau>\tau_1>0>\tau_2)\\
&&\theta_6\equiv \theta(\tau_1>\tau>\tau_2>0)\\
&&\theta_0\equiv \theta(\tau>0)\theta(\tau_1>\tau_2)
\eea
The integration over these ranges appear 	with a integrand that is only a function of $\tau_1-\tau_2$. Denoting,
\be
I_i\equiv\int{d\tau_1d\tau_2\theta_i}G(\tau_1-\tau_2)
\ee
we have
\bea
I_1&=&\int_0^{\beta/2-\tau}d\Delta\tau[\beta/2-\tau-\Delta\tau]G(\Delta\tau)\\
I_2&=&\int_0^{\beta/2}[\beta/2-\Delta\tau]G(\Delta\tau)\\
I_3&=&\int_0^\tau d\Delta\tau [\tau-\Delta\tau]G(\Delta\tau)\\
I_4&=&\int_{\tau}^{\beta/2}d\Delta\tau[\Delta\tau-\tau]G(\Delta\tau)\\
&&\quad+\int_{\beta/2}^{\tau+\beta/2}d\Delta\tau [\beta/2-\tau]G(\Delta\tau)\\
&&\quad+\int_{\tau+\beta/2}^{\beta}d\Delta\tau[\beta-\Delta\tau]G(\Delta\tau)
\eea
$I_5$ and $I_6$ sometimes appear with an integrand that depends on both $\Delta\tau$ and $\bar\tau$. In that case,
\bea
I_5&=&\int_0^\tau d\Delta\tau\int_{-\Delta\tau/2}^{\Delta\tau/2}\\
&&+\int_\tau^{\beta/2}{d\Delta\tau/2}\int_{-\Delta\tau/2}^{\tau-\Delta\tau/2}d\bar\tau\\
&&+\int_{\beta/2}^{\tau+\beta/2}d\Delta\tau\int_{-\beta/2+\Delta\tau/2}^{\Delta\tau/2}d\bar\tau\\
I_6&=&\int_0^\tau d\Delta\tau\int_{\tau-\Delta\tau/2}^{\tau+\Delta\tau/2}d\bar\tau\\
&&+\int_\tau^{\beta/2-\tau}d\Delta\tau\int_{\Delta\tau/2}^{\tau+\Delta\tau/2}d\bar\tau\\
&&+\int_{\beta/2-\tau}^{\beta/2}d\Delta\tau\int_{\Delta\tau/2}^{\beta/2-\Delta\tau/2}d\bar\tau
\eea
If the integrand only depends on $\tau_1-\tau_2$, we find
\bea
I_5&=&\int_0^\tau d\Delta\tau(\Delta\tau)G(\Delta\tau)\\
&&+\tau\int_0^{\beta/2}d\Delta\tau G(\Delta\tau)\\
&&+\int_{\beta/2}^{\beta/2+\tau}d\Delta\tau(\tau+\beta/2-\Delta\tau)G(\Delta\tau)\\
I_6&=&\int_0^\tau d\Delta\tau (\Delta\tau)G(\Delta\tau)\\
&&+\tau\int_\tau^{\beta/2-\tau}d\Delta\tau G(\Delta\tau)\\
&&+\int_{\beta/2-\tau}^{\beta/2}d\Delta\tau (\beta/2-\Delta\tau)G(\Delta\tau)
\eea
The $\tau=0$ boundary in these integrals has to be replaced by the inverse UV cutoff. Moreover, since we use the expression of $D_d(\tau)\sim 1/\tau^{d-1}$ defined for $\tau\in(-\beta/2,\beta/2)$, the integrals have to be folded back to this range using periodicity of the Green's functions. The final result is
\bea
{\cal M}_\perp(\tau)&=&\frac{4}{3}+\frac{2}{9}u\beta+\gamma^2\Big[\frac{1}{9}\beta f(0)-\frac{4\tilde S_{d+1}}{3}\frac{\tau^\eps}{\eps}\Big]\nonumber\\
&&+\frac{\gamma^2u}{3}\beta\Big\{\Big[\frac{1}{9}-\frac{2\tau}{\beta}+\frac{\tau^2}{\beta^2}\Big]\beta f(0)-\frac{8\tilde S_{d+1}}{9}\frac{I_\perp}{\eps}\Big\}\nonumber\\
&&+\frac{u^2\beta^2}{3}\Big[\frac{1}{9}+\frac{\tau^2}{\beta^2}-\frac{2\tau}{\beta}\Big]
\eea
where $I_\perp=(3/2)[(\beta/2)^\eps-\tau^\eps/3]=1+{\cal O}(\eps)$ and we have defined
\be
f(\mu_0^{-1})=\Big[\int_{\mu_0^{-1}}^{\beta/2}+\int_{-\beta/2}^{-\mu_0^{-1}}\Big]{d\tau D_{3-\eps}(\tau)}=2\tilde S_{d+1}\mu_0^{1-\eps}.
\ee
That this is independent of the IR cut-off is an artefact of using the $T=0$ expression of $D(\tau)$. We are only interested in the UV-part of this expression and the IR-dependence is not important. Similarly,
\bea
{\cal M}_z(\tau)&=&\frac{2}{3}-\frac{2}{9}u\beta-\gamma^2\Big[\frac{1}{9}\beta f(0)+\frac{4\tilde S_{d+1}}{3}\frac{\tau^\eps}{\eps}\Big]\nonumber\\
&&+{\gamma^2u}\beta\Big\{-\frac{1}{27}\beta f(0)+\frac{10\tilde S_{d+1}}{9}\frac{I_z}{\eps}\Big\}\nonumber\\
&&-\frac{1}{27}u^2\beta^2
\eea
where $I_z=(3/5)[2(\beta/2)^\eps-\tau^\eps/3]=1+{\cal O}(\eps)$, and
\bea
{\cal M}_{u}(\tau)&=&\frac{2}{9}+\frac{2}{27}u\beta+\gamma^2\Big[\frac{1}{27}\beta f(0)-\frac{4\tilde S_{d+1}}{3}\frac{\tau^\eps}{\eps}\Big]\nonumber\\
&&-{\gamma^2u}\beta\Big\{\frac{1}{27}\beta f(0)+\frac{2\tilde S_{d+1}}{9}\frac{\tau^\eps}{\eps}\Big\}\nonumber\\
&&-\frac{1}{27}u^2\beta^2
\eea
These correlation functions diverge in two ways: one as $\eps\to 0$ or $d\to 3^-$. And the other one through the explicit UV cut-off dependence $\mu_0\to\infty$. The goal is to renormalize the coupling constants to remove this divergences. \\

\subsection{Renormalization}
As discussed in \cite{Whitsitt17}, the bulk renormalization is achived by
\be
\phi_\alpha=\sqrt{Z}\tilde \phi_\alpha, \quad g_0=\frac{\mu^\eps Z_g}{S_{d+1}Z^2}g, \quad S_d=\frac{2}{\Gamma(d/2)(4\pi)^{d/2}}
\ee
where
\bea
Z&=&1-\frac{10}{144}\frac{g^2}{\eps},\\
Z_g&=&1+\frac{5}{3}\frac{g}{\eps}+\frac{}{}\frac{g^2}{\eps^2}-\frac{}{}\frac{g^2}{\eps},
\eea
leading to the beta function [$d\ell\equiv-d\log\mu$]
\be
\beta_g\equiv \frac{dg}{d\ell}=\eps g-\frac{5}{3}g^2,
\ee
which has a fixed point at $g^*=3\eps/5$.

In order to renormalize the impurity problem, we define the renormalized (tilde) coupling constants as
\be
\gamma= \tilde\gamma \mu^{\eps/2}A_\gamma, \quad A_\gamma=\frac{Z_\gamma}{\sqrt{\tilde S_{d+1}Z_\perp Z}}, \quad u= u'+Z_u \tilde u\mu\label{eqZ}
\ee
$u'$ is introduced to absorb the non-universal part of the $u$. We assume $\beta\mu=cte$ is a constant that can be absorbed into a redefinition of $\tilde u$. This implies we are comparing theories in which the temperature is equal to the energy scale of interest $T\sim\mu$. The renormalized (tilde) correlation functions as
\be
{\cal M}_\perp(\tau) = Z_\perp\tilde {\cal M}_\perp(\tau), \qquad {\cal M}_z(\tau)=Z_z\tilde M_z(\tau)
\ee 
and
\be
{\cal M}_{u}(\tau)=Z_u^2\tilde{\cal M}_{u}(\tau).
\ee
We see that first, the non-universal UV dependence can be removed by choosing 
\be
u'=-\frac{\gamma^2}{2} [f(\mu_0^{-1})-f(\mu^{-1})]={\gamma^2}\tilde S_{d+1}[\mu^{1-\eps}-\mu_0^{1-\eps}],
\ee
and second, the $1/\eps$ divergence can be eliminated by choosing
\bea
Z_u&=&1-\frac{3\tilde\gamma^2}{\eps}\\
Z_\perp&=&1-\frac{\tilde\gamma^2}{\eps}-\tilde u\frac{\tilde\gamma^2}{\eps}\\
Z_z&=&1-2\frac{\tilde\gamma^2}{\eps}+2\tilde u\frac{\tilde\gamma^2}{\eps}
\eea
Taking derivative of Eqs.\,\pref{eqZ} w.r.t $d\ell=-d\log\mu$, we find
\bean
&&\Big[1+\tilde\gamma\frac{d\log A_\gamma}{d\gamma}\Big]\beta_\gamma+\tilde\gamma\frac{d\log A_\gamma}{du}\beta _u+\tilde\gamma\frac{d\log A_\gamma}{dg}\beta_g=\frac{\eps}{2}\tilde\gamma\\
&&\tilde u\frac{d\log Z_u}{d\gamma}\beta_\gamma+\Big[1+\tilde u\frac{d\log Z_u}{du}\Big]\beta_u+\tilde\gamma\frac{d\log Z_u}{dg}\beta_g=\tilde u+\mu\frac{du'}{d\mu}
\eean
 Ref.\,\cite{Whitsitt17} also calculate vertex corrections by the bosonic interaction, leading for $S=1$ to
\be
Z_\gamma=1+\frac{2\pi^2}{9}\frac{g\gamma^2}{\eps}
\ee
It can be shown that $Z$ and $Z_\gamma$ are not important to ${\cal O}(\eps)$ and we can use the simplification $A_\gamma\propto Z_\perp^{-1/2}$. Moreover, the bulk interaction $g$ will not play a role to ${\cal O}(\eps)$. Inverting the resulting matrix
\bea
&&\mat{1-(\tilde\gamma/2)\partial_\gamma \log Z_\perp & -(\tilde\gamma/2)\partial_u\log Z_\perp  \\
\partial_\gamma\log Z_u & 1+\tilde u\partial_u\log Z_u
}\mat{\beta_\gamma \\ \beta_u}\qquad\qquad\nonumber\\
&&\hspace{5cm}=\mat{(\eps/2)\tilde\gamma \\ \tilde u+Z_u^{-1}\partial_\mu u'},
\eea
we find the beta functions reported in Eq.\,\pref{eqbeta} of the main text.

Since $\tilde{\cal M}_\perp(\tau)\sim (\mu\tau)^{-\eta_\perp}$ and the bare correlation function ${\cal M}_\perp(\tau)=Z_\perp \tilde {\cal M}(\tau)$ is independent of $\mu$, we find
\bean
-\eta_\perp&=&\frac{dZ_\perp}{d\log\mu}=\frac{d\log Z_\perp}{d\gamma}\beta_\perp+\frac{d\log Z_\perp}{du}\beta_u+\frac{d\log Z_\perp}{dg}\beta_g\\
-\eta_z&=&\frac{dZ_z}{d\log\mu}=\frac{d\log Z_z}{d\gamma}\beta_\perp+\frac{d\log Z_z}{du}\beta_u+\frac{d\log Z_z}{dg}\beta_g\\
-\eta_u&=&\frac{dZ_u^2}{d\log\mu}=\frac{d\log Z_u^2}{d\gamma}\beta_\perp+\frac{d\log Z_u^2}{du}\beta_u+\frac{d\log Z_u^2}{dg}\beta_g
\eean
Again, to ${\cal O}(\eps)$ we can drop the bulk contribution in the last terms.  This gives the exponents $\eta_\perp=\gamma^2$, $\eta_z=2\gamma^2$ and $\eta_u=6\gamma^2$ discussed in the paper.
\subsection{Relation between $\eta$ and $\nu$}
Here, we discuss the relation between the exponent in the correlation function and the free energy. For example, from
\be
\braket{T_\tau S_z(\tau)S_z}\sim \tau^{-\eta_z}
\ee
the corresponding susceptibility is found to be
\be
\chi_z=\intob{d\tau }\braket{T_\tau S_z(\tau)S_z}=\beta^{1-\eta_z}. \label{eqeta}
\ee
This susceptibility can be obtained by using taking derivative w.r.t. a source term $h_zS_z$ in the action. In presence of the source term, the scaling form of the free energy is
\be
F(h_z)=b^{-1}\Phi(h_zb^{1/\nu_z}).
\ee
where $b\sim\beta$ is a given length-scale. We find
\be
\chi_z=\frac{d^2F}{dh_z^2}=\beta^{2/\nu_z-1}\label{eqnu}
\ee
By comparing Eqs.\,\pref{eqeta} and \pref{eqnu} we find
\be
\nu_z=[1-\eta_z/2]^{-1}.
\ee
\subsection{Lattice Model}
Simulations by worm algorithm allow us to perform a comprehensive study of the universal properties of an impurity in a two-dimensional $O(2)$ quantum critical environment. As long as we are interested in the critical properties only, we are allowed to simulate the environment system with a J-current model with a spin impurity at the origin. 

In the $d=2+1=3$ case, the bulk part of this model consists of integer currents $J$ living on the bonds of a three dimensional $L^2\times L_{\tau}$ cubic lattice, with $L$ as the size of the spatial dimensions and $L_\tau$ as the size along the ``temporal" direction (in the absence of the impurity, all the three dimensions are absolutely equivalent). The currents are subject to the zero-divergence constraint, 
\begin{equation}
{\rm div}\, J \, =\, 0
\label{divJ} ,
\end{equation}
meaning that at each site, the algebraic---incoming minus outgoing---sum of all the currents is zero. To have a really minimalistic model, one
also confines the allowed values of the bond currents to just three numbers:
\begin{equation}
J \, =\, 0, \, \pm 1.
\label{J_values} 
\end{equation}
The Hamiltonian of the model reads
\begin{equation}
\label{J-current}
 H_J \, =\,  \frac{1}{2K}  \sum_{i, \hat{e}} J_{i, \hat{e}}^2  \qquad \quad (\hat{e} = \hat{x}, \hat{y}, \hat{\tau}).
\end{equation}
Here the vector $i=(x,y,\tau)$ labels the sites on the cubic lattice by three discrete coordinates: $x$, $y$, and $z$; $\hat{x}$, $\hat{y}$, 
and $\hat{\tau}$ are the lattice unit translation vectors in corresponding directions;  $J_{i, \hat{e}} \equiv -J_{i+\hat{e}, -\hat{e}}$ is the J-current of the bond going from the site $i$ in the direction $\hat{e}$.

In terms of the mapping onto a two-dimensional system of lattice bosons (at an integer filling factor), the closed loops of currents should be understood as the worldlines of O(2) charge quanta, with $J_{i=(x,y,\tau), \hat{\tau}}$ having the meaning of the particle/hole charge on the site
$(x,y)$ at the imaginary-time moment $\tau$. This model (\ref{divJ})--(\ref{J-current}) describes the universal properties of the insulator-to-superfluid criticality; the corresponding transition takes place at the critical value $K_c = 0.3332052(20)$~\cite{huang2016trapping} of the control parameter  $K$. 

We now discuss the implementation of the spin impurity in the J-current models. The model used in this paper is inspired by the spin-$1/2$ impurity model for the halon effect ~\citep{chen2018halon}. We introduce a spin-$1$ degree of freedom at the origin by replacing the original J currents with the spin currents on the bonds going from the sites $(0,0,\tau)$ in the direction $\hat{\tau}$. For spin-$1$ impurity, the spin current can only takes three values,
\begin{equation}
S_{\tau} = 0, \pm{1} .
 \label{half-current}
\end{equation}
At the impurity site, the zero-divergency condition also includes the algebraic sum of the spin currents associated with this site, which guarantees the conservation of total charge: the $\tau$-independence of the total charge number $Q$, where
\begin{equation}
 Q \, =\, S_{\tau} + \sum_{x,y} J_{(x,y,\tau), \hat{\tau}} .
\label{J_charge_S}
\end{equation}

The impurity Hamiltonian contains an interaction term and a possible magnetic field term:
\begin{equation}
\label{Jcurrent-imp}
 H_{\rm imp} \, =\frac{U}{2} \sum_{\tau} S_{\tau} ^2+\,h_z \sum_{\tau} S_{\tau} .
\end{equation}
A bare spin-$1$ with three degenerate states corresponds to the parameter $U=0$. In our simulations, we turn off the magnetic field term which explicitly breaks the $Z_2$ symmetry of the Hamiltonian.

The coupling between the impurity and the environment is introduced as:
\begin{equation} 
\label{Jcurrent-imp2}
H_{\rm imp-bulk} = {1\over 2K_{I}} J_{(0,0,\tau),\hat{e}} ^2  \qquad \quad (\hat{e} = \hat{x}, \hat{y}).
\end{equation}
When the bulk is fine-tuned to the critical point $K=K_c$, the universal physics of the impurity model should be independent of the choice of $K_I$. Therefore, we fix $K_I=K$ for simplicity.

We also point out that the above lattice model with a spin-$1$ impurity can be easily generalized to generic spin-$S$ impurity by allowing the spin current to fluctuate between $-S, -S+1, ..., S-1, S$.

\end{document}